\documentclass{article}
\usepackage[english]{babel}
\usepackage[utf8x]{inputenc}
\usepackage{bm}
\usepackage{amsmath, amssymb}
\usepackage{polski}
\usepackage{graphicx}

\usepackage{natbib}
\setcitestyle{authoryear,open={(},close={)}}

\title{The hierarchical generalized linear model and the bootstrap estimator of the error of prediction of loss reserves in a non-life insurance company}
\author{Alicja Wolny-Dominiak}

\begin{document}
\maketitle

\begin{abstract}
This paper presents the hierarchical generalized linear model (HGLM) for loss reserving in a non-life insurance company. Because in this case the error of prediction is expressed by a complex analytical formula, the error bootstrap estimator is proposed instead. Moreover, the bootstrap procedure is used to obtain full information about the error by applying quantiles of the absolute prediction error. The full \textbf{R} code is available on the Github \texttt{https://github.com/woali/BootErrorLossReserveHGLM}.
\end{abstract}

\smallskip
\noindent \textbf{Keywords.} Loss Reserves; Hierarchical Generalized Linear Model; Error of prediction; MSE; Parametric Bootstrap\\

JEL: G22, C21, C53

\section{Introduction}
This analysis concerns a non-life insurance company. The largest item on an insurer's balance sheet are technical provisions. Any variations in their values have a great impact on the insurer's financial strength. A large part of the provisions is the reserve for incurred but not reported (IBNR) claims or simply – the loss reserve, which is crucial to the insurer’s solvency. The total loss reserve is generally determined by statistical methods based on both deterministic techniques and stochastic models and is a sum of outstanding claims liabilities.

A wide variety of the total loss reserve prediction tools are presented in literature in the group of stochastic methods. For years the most popular method with insurers was the chain-leader technique (e.g. \citep{mack1994stochastic}, \citep{mack2000comparison}, \citep{verrall2000investigation}). Parallel methods based on the regression model have been developed (e.g. \citep{kremer1982ibnr}, \citep{mack1991simple}). In \citep{renshaw1998stochastic} it is shown that the chain-leader method is equivalent to the well-known generalized linear model (GLM) with fixed effects assuming over-dispersed Poisson (ODP) distribution for incremental claims. This paper puts forward a mixed model for loss reserving as an extension of the GLM in which data for different homogenous groups are assumed to be dependent in some way. The model belongs to the class of HGLMs and contains both fixed effects as well as random effects. Model's parameters are estimated in the frequentist approach using extended likelihood \citep{lee1996hierarchical}, \citep{lee2001hierarchical}, \citep{lee2003extended}. As the total loss reserve is actually a random variable, its value has to be predicted. In order to measure the prediction accuracy, the bootstrap root mean squared error of prediction (RMSEP) estimator is proposed.

\section{Data -- single loss triangle and total loss reserve}
In loss reserving, data have a specific form referred to as the loss triangle. Consider the random variable $Y_{ij}$ with $y_{ij}$ realizations $i,j=0,\ldots,n$ being the incremental quantity for insurance claims that occurred in year $i$ (origin year) and were reported to the insurer after $j$ years (development year). It could be the value of claims as well as their number. The matrix $[Y_{ij}]_{(n+1) \times (n+1)}$ represents random variables in the loss triangle for a single LOB as shown in Table 1 below.

\begin{table}[!hbtp]
\centering
\small
\caption{Incremental loss triangle for $k$th group \label{tr}}
\begin{tabular}{c||ccccc}
  \hline
 i/j &0 & ... & $n-1$ & $n$ & \\ \hline \hline
0& $Y_{00}$ & ... & $Y_{0n-1}$ & $Y_{0n}$ &\\ \hline
1& $Y_{10}$ & ... & $Y_{1n-1}$ &  & \\ \hline
...& ...  & ... & &  & \\ \hline
n& $Y_{n0}$ &  &  &  & \\ \hline
\end{tabular}
\end{table}

The realizations $y_{ij}$ for $i+j \leq n$ are observed data while $y_{ij}$ for $i+j>n$ represent the future unobserved data. The goal in loss reserving is to predict the random variable $R$ being the total loss reserve defined as the sum of unknown future claims:
\begin{equation}\label{R}
R=\sum_{i,j;i+j>n}Y_{ij}
\end{equation}
%assuming that $Y_{00},...,Y_{nn}$ are independent.
In order to asses the uncertainty of predictor $\hat R$, usually
\begin{equation}
RMSEP(\hat R)=\sqrt{E[(\hat R-R)^2]}
\end{equation}
is adopted.

\section{The model and the prediction}
Let us assume $Y_{ij}\sim ODP(\mu_{ij}, \phi)$ as independent and identically distributed (i.i.d) with $E[Y_{ij}]=\mu_{ij}$ and $Var(Y_{ij})=\phi\mu_{ij}$ and the unknown dispersion parameter $\phi$. In the GLM the mean has a linear form $\log(\mu_{ij})=\bm{x}_{ijk}'\bm{\beta}$, where $\bm{\beta}=(c,u_1,\ldots,u_n,\beta_1,\ldots,\beta_n)'$ is the vector of fixed effects, $\bm{x}_{ijk}'$ is the row of the design matrix $\bm{X}$ and $u_0=\beta_0=1$. The dispersion parameter $\phi$ is the same for all elements of the loss triangle. It is easy to extend this model to the general class of models assuming that incremental claims follow the three-parametric Tweedie $T(\mu, \phi, p)$ \citep{jorgensen1987exponential} with dispersion parameter $\phi$ and power $p\in (0,\infty ]\setminus (0,1)$. The GLM is then of the form:
\begin{equation}\label{glm}
E[Y_{ij}]=\mu_{ij}=\exp(\bm{x}_{ij}'\bm{\beta}), Var(Y_{ij})=\phi\mu_{ij}^p,
\end{equation}
In order to draw an inference about the total loss reserve $R$, MLE estimators of parameters $(\hat c, \hat u_1,\ldots,\hat u_n,\hat \beta_1,\ldots,\hat \beta_n,\hat \phi)'$ can be plugged into Formula (\ref{R}) giving the predictor:
\begin{equation}\label{Rglm}
\hat R_{GLM}=\sum_{i,j;i+j>n} \hat Y_{ij}=\sum_{i,j;i+j>n} \exp(\hat c+\hat u_i+\hat \beta_j).
\end{equation}
The total loss reserve prediction accuracy is typically measured using the $RMSEP$. According to \citep{england1999analytic}, the predictor defined in Formula (\ref{Rglm}) can be expressed as:
\begin{align}\nonumber
&RMSEP(\hat R_{GLM})=\sum_{i,j;i+j>n}\phi_{ij}\mu_{ij}^p+\sum_{i,j;i+j>n} \mu_{ij}^2Var(\bm{x}_{ij}'\bm{\beta})+\\
&+\sum_{i,j;i+j>n}\mu_{j_1j_1}\mu_{j_2j_2}Cov(\bm{x}_{i_1j_1}'\bm{\beta},\bm{x}_{i_2j_2}'\bm{\beta}).
\end{align}

The fundamental assumption in the GLM is the independence between random incremental claims $Y_{ij}$, $i,j=0,\ldots,n$ taken from a single loss triangle. However, this assumption may be inadequate, as indicated by the fact that origin years reflect the same process of loss development. This means that there is a dependence between claims from the same development year $j$ but from different origin years. The dependence may be taken into account by applying a mixed model with fixed and random effects in place of the GLM with fixed effects only.

Let us assume that effects $u_0,\ldots,u_n$ occuring in model (\ref{glm}) are independent realizations of random variable $U$ following Tweedie distributions $u_i\sim T(\psi_{iu},\phi_{u})$ with $E[u_i]=\psi_{iu}$ and $Var(u_i)=\phi_{u}\psi_{iu}^p$. This means that incremental claims from the loss triangle are now conditionally independent, which can be described in the following way: $Y_{ij}$ of different origin years are independent, but due to the random effects, for any origin year $i$, the random variables $Y_{ij}$ for the different development years $j$ are dependent (cf. \citep{gigante2013prediction}, pp. 383, (a1)--(4)). This approach is examined in \citep{gigante2013claims}, \citep{gigante2013prediction}, where the hierarchical GLM (HGLM) is described. The authors assume a conditional ODP distribution for incremental claims  $Y_{ij}|u_i$. Like for the GLM, this assumption is extended to any $p$ such that $Y_{ij}|u_i\sim T(\mu_{ij},\phi,p)$, which gives the following general form of the HGLM:
\begin{align}\label{hglm}\nonumber
&E[Y_{ij}|u_i]=\mu_{ij}(u_i)=\exp(\bm{x}_{ij}'\bm{\beta}+\bm{z}_{ij}'\bm{v})\\
& Var(Y_{ij}|u_i)=\phi\mu_{ij}^p(u_i),
\end{align}
where $\bm \beta=(c, \beta_1,\ldots,\beta_n)'$, $v=(\log(u_1),\ldots,\log(u_n))'$ and $\bm{z}_{ijk}$ is the row of the design matrix $\bm{Z}$. Under the model defined in (\ref{hglm}) and using the conditional independence assumption, the total loss reserve is defined as the conditional expected value:
\begin{equation}\label{expval}
R_{HGLM}=\sum_{i,j;i+j>n} E[Y_{ij}|u_i]=\sum_{i,j;i+j>n} \exp(\bm{x}_{ij}'\bm{\beta}+\bm{z}_{ij}'\bm{v})
\end{equation}
The estimation of the HGLM parameters is more complex than in the case of the GLM due to the presence of random effects. Following \citep{gigante2013claims}, the likelihood-based approach is applied using the h-likelihood
function. The idea is to treat the vector of random effects $\bm v$ as the vector of fixed effects and transform the HGLM into an augmented GLM. Details concerning the h-likelihood estimation are presented in two fundamental papers \citep{lee1996hierarchical}, \citep{lee2001hierarchical}. This numerical analysis uses the $\bm R$ implementation of the algorithm taken from the \textit{hglm} package \citep{hglmR}. After estimators $\hat {\bm\beta}$ and predictors $\hat{ \bm v}$ are found, the total loss reserve predictor is of the form:
\begin{equation}\label{expval1}
\hat R_{HGLM}=\sum_{i,j;i+j>n}\exp(\hat c+\hat \beta_j)\hat u_i
\end{equation}
A similar mixed model-based approach is described in \citep{antonio2006lognormal}, but only for a lognormal distribution. The authors adopted the generalized linear mixed model (GLMM) and the REML as the method of estimation of the model parameters.

\section{Bootstrap MSEP}
In \citep{gigante2013prediction} the estimator of $RMSEP(\hat R)$ is derived analytically in the case of a single loss triangle. Unfortunately, the formula describing the estimator has a rather complex structure. An alternative may be to use the parametric bootstrap technique, like for the $RMSEP(\hat R_{GLM})$ proposed in \citep{england1999analytic}. This paper proposes another procedure.

Let us denote the realization of $Y_{ij}$ as $y_{ij}$. As the HGLM is commonly an extension of the GLM, Pearson residuals $r_{ij}=\frac{\hat y_{ij}-\mu_{ij}}{\sqrt{\mu_{ij}}}$ are bootstrapped instead of the data from the loss triangles. A single $b$th simulation, $b=1,\ldots,B$ in the full bootstrap procedure has the following steps:
\begin{itemize}
\item{take a sample $r_{ij}^{*b}$ with replacement from Pearson residuals $r_p$ based on data from the loss triangle $i+j \leq n$}
\item{calculate new data $y_{ij}^{*b}=r_{ij}^{*b}\sqrt{\hat{\mu}_{ij}}+\hat{\mu}_{ij}$, $i+j \leq n$}
\item{estimate the vector of parameters $(\bm \beta^{*b}, \bm{v}^{*b}, \phi^{*b}, \phi_{u}^{*b})'$ based on new data $y_{ijk}^{*b}$}
\item{calculate predicted values $\hat y_{ij}^{*b}=\exp(\hat c + \hat \beta_j)\hat u_i $ for $i+j>n$}
%\item{generate values $u_1^{*b},\ldots,u_n^{*b}$ from Tweedie distribution with $\phi_u^{*b}$}
\item{generate values $y_{ij}^{*b}$ for $i+j>n$ from Tweedie distribution with the dispersion parameter $\phi^{*b}$}
%\item{calculate the predictor of the total loss reserve as:
%\begin{equation}
%\hat R^{*b}=\sum_{j=n-i+1}^n\hat{y}_{ijk}^{*b}
%\end{equation}}
\end{itemize}

After $B$ iterations of the bootstrap procedure the estimator of $RMSEP(\hat R_{HGLM})$ is of the following form:
\begin{equation}\label{brmsep}
\widehat{RMSE}(\hat R)=\sqrt{\frac{1}{B}\sum_{b=1}^B(\sum_{i,j;i+j>n}\hat{y}_{ij}^{*b}-\sum_{i,j;i+j>n}{y}_{ij}^{*b})^2}
\end{equation}

Let us notice that the commonly used $RMSEP$ gives information only about the average error of prediction. If full distribution of the prediction error is of interest, quantiles of the absolute prediction error can be used (see \citep{zadlo}), where the $p$th quantile is defined as $Q_p(\hat R)=Q_p(|\hat R-R|)$. Quantiles reflect the relation between the magnitude of the error and the probability of its realization. In order to estimate $Q_p(\hat R)$, the bootstrap procedure described above can be applied. In the last step, Formula (\ref{brmsep}) is replaced by the formula presented below:
\begin{equation}\label{qp}
\widehat{Q_p}(\hat R)=Q_p(|\sum_{i,j;i+j>n}\hat{y}_{ijk}^{*b}-\sum_{i,j;i+j>n}{y}_{ijk}^{*b}|)
\end{equation}
As a result, the bootstrap estimator of the $Q_p(\hat R)$ error is obtained.

\section{Numerical example}
The subject of this analysis is the single loss triangle taken from \citep{wuthrich2008stochastic}, pp. 33, like in \citep{gigante2013claims}. In order to demonstrate the HGLM for loss reserving, it is assumed that incremental claims conditionally follow Tweedie distribution $Y_{ij}|u_i\sim T(\mu_{ij},\phi_u)$ with power $p=1$ and random effects follow Tweedie distribution $u_i \sim T(\psi_u, \phi_u)$ with power $p=2$. This model is equivalent to the ODP-Gamma HGLM. Moreover, dispersion parameters $\phi$, $\phi_u$ are assumed as constant for both incremental claims and random effects, respectively. The vector of parameters
$$(c, \beta_1, \ldots ,\beta_9,u_0, \ldots ,u_9, \phi, \phi_u)'$$
is estimated using the h-likelihood function. Plugging the obtained values into Formula (\ref{expval1}), the total loss reserve and the loss reserves for origin years are obtained. In order to find the estimator of $RMSEP$, the bootstrap procedure was implemented. The number of simulations was $n=1000$. The results are presented in Table \ref{t}.

\begin{table}[!hbtp]
\centering
\small
\caption{Total loss reserve and bootsrap estimator $\widehat{RMSE}$}
    \begin{tabular}{ccccccc}
    \hline
    Origin year $i$ & $\hat R_{HGLM}$ & $\widehat {RMSEP}_{HGLM}$ & $\hat R_{GLM}$ & $\widehat {RMSEP}_{GLM}$ \\ \hline
  1     & 15 239 & 19 763 & 15 125 & 19 620 \\
    2     & 26 415 & 24 220 & 26 257 & 24 508 \\
    3     & 35 264 & 27 182 & 34 538 & 27 613 \\
    4     & 86 572 & 38 331 & 85 302 & 40 690 \\
    5     & 157 840 & 49 339 & 156 494 & 54 876 \\
    6     & 290 991 & 66 545 & 286 121 & 70 027 \\
    7     & 465 024 & 81 939 & 449 161 & 86 568 \\
    8     & 1 078 769 & 130 374 & 1 043 237 & 132 874 \\
    9     & 3 988 971 & 281 813 & 3 950 809 & 315 437 \\ \hline
    Total & 6 145 085 & 373 914 & 6 047 044 & 403 506 \\ \hline
     & $\phi$=14 739 & &$\phi$=14 714 & \\
    & $\phi_u$=0.0054 & & \\ \hline
    \end{tabular}%
  \label{t}%
\end{table}%

It can be seen that the HGLM gives higher loss reserve values compared to the GLM but the errors are generally lower. Naturally, errors rise for subsequent origin years in both cases, which results directly from the fact that there are fewer and fewer observations in the loss triangle. The two models are not compared directly because in the case of the HGLM it is the conditional value of reserves that is determined.

\newpage
\begin{figure}[htbp]
\centering
\includegraphics[scale=0.4]{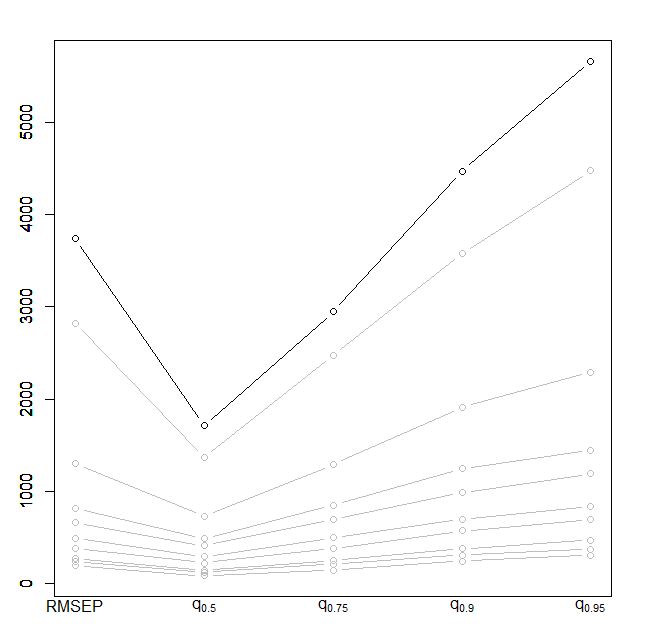}
\caption{Measures of accuracy for origin years and the total loss reserve\label{q}}
\end{figure}

\newpage
The bootstrap procedure makes it possible to obtain not only the RMSEP but also the quantiles of the absolute error of prediction $Q_p$. Figure \ref{q} presents quantiles $Q_{0.5}$, $Q_{0.75}$, $Q_{0.9}$ and $Q_{0.95}$ for origin years and for the total loss reserve. It can be seen that in earlier origin years the RMSEP value is more or less equal to quantiles of the order of 0.75, whereas for the total loss reserve the value is close to quantiles of the order of 0.9.

\section{Conclusions}
GLMs are popular statistical techniques in actuarial practice, especially in ratemaking but also in loss reserving. However, the independence assumption needed in GLMs is generally violated in many cases. There are three basic advantages of the mixed HGLM application. Firstly, by introducing random effects into the model, the dependencies between development years are taken into account. Secondly, by imposing certain constraints on random effects, it is possible to take account of external information which does not come from the sample directly, like in \citep{gigante2013claims}, and which has an impact on the total loss reserve value. Thirdly, distributions can be set flexibly within the Tweedie family, e.g. the Gamma, the inverse-Gaussian or the compound Poisson distribution. The downside, however, is the complex form of the error prediction. Therefore, it is proposed herein that the error should be determined by means of the bootstrap technique. Although this solution is not perfect, its important advantage is that full information on the absolute error distribution can be obtained easily using quantiles.

\bibliographystyle{abbrvnat}
\bibliography{rezerwy}

\end{document}